\documentclass[aps,pra,superscriptaddress,amsmath,amssymb,preprintnumbers,showpacs,twocolumn]{revtex4}
\usepackage{amssymb}
\usepackage{graphicx}
\usepackage{bm}
\usepackage{color}
\usepackage{subfigure}

\newcommand{\Ket}[1]{\left\vert #1\right\rangle}
\newcommand{\Bra}[1]{\left\langle #1\right\vert}
\newcommand{\SimKet}[1]{\vert #1\rangle}
\newcommand{\SimBra}[1]{\langle #1\vert}

\newcommand{\KetBra}[2]{\left\vert#1\right\rangle\left\langle#2\right\vert}
\newcommand{\Projector}[1]{\KetBra{#1}{#1}}

\renewcommand{\eqref}[1]{(\ref{#1})} %{Eq.~(\ref{#1})}

\newcommand{\trace}{\mathrm{tr}}

\begin{document}

\title{Genuine Tripartite Entanglement in a Spin-Star Network at Thermal Equilibrium}

\author{B. Militello}
\email{bdmilite@fisica.unipa.it} \affiliation{Dipartimento di
Fisica dell'Universit\`{a} di Palermo, Via Archirafi 36, 90123
Palermo, Italy}

\author{A. Messina}
\affiliation{Dipartimento di Fisica dell'Universit\`{a} di
Palermo, Via Archirafi 36, 90123 Palermo, Italy}

\begin{abstract}
In a recent paper [M. Huber {\it et al}, Phys. Rev. Lett. {\bf
104}, 210501 (2010)] new criteria to find out the presence of
multipartite entanglement have been given. We exploit these tools
in order to study thermal entanglement in a spin-star network made
of three peripheral spins interacting with a central one. Genuine
tripartite entanglement is found in a wide range of the relevant
parameters. A comparison between predictions based on the new
criteria and on the tripartite negativity is also given.
\end{abstract}


%Entanglement and quantum nonlocality;
%Entanglement measures, witnesses, and other characterizations;
%Quantized spin models, including quantum spin frustration;

\maketitle

\section{Introduction}

Entanglement has been widely studied for decades: criteria to find
out the presence of bipartite entanglement in a quantum state are
well known \cite{ref:Revealing_Bipartite}, and, for systems with
few degrees of freedom, it can be quantified \cite{ref:Wootters},
whether the relevant state is pure or mixed. The analysis of
multipartite entanglement is a more complicated task. For example,
there have been many proposals of tripartite entanglement
quantifiers
\cite{ref:TripQuantify-1,ref:TripQuantify-2,ref:TripQuantify-3,ref:TripQuantify-4}
and witnesses \cite{ref:TripWitnesses-1, ref:TripWitnesses-2,
ref:TripWitnesses-3}, but none of such contributions have given a
definitive solution to the problem of singling out and quantifying
this type of correlations \cite{ref:Fazio2008}. The three-tangle
has been considered a good tool able to quantify tripartite
entanglement in pure states \cite{ref:TripQuantify-1}, but
recently it has been criticized \cite{ref:DoesThreeTangle}.
Difficulties grow up when the system is described by a mixed
state. Indeed, many of the proposals previously mentioned are
valid only for pure states. An interesting tool for detecting
tripartite correlations in mixed states has been presented by
Sabin and Garcia-Alcaine \cite{ref:Sabin2008}, but the tripartite
negativity they introduced (i.e., the geometric mean of the three
negativities associated to the three possible bipartitions of a
tripartite system) is not able to tell a genuine tripartite
entangled state from a state which is biseparable in a generalized
sense. Very recently, Huber {\it et al} \cite{ref:Huber2010} have
given a set of relations that provide sufficient conditions to
assert the presence of multipartite entanglement in an
indisputable way, whether the state under scrutiny is pure or
mixed. The basic idea of such criteria is to exclude the presence
of any form of biseparability, in connection with all the possible
bipartitions.

Over the last decade, the concept of thermal entanglement has
emerged by investigating the presence of quantum correlations in
quantum systems at thermal equilibrium \cite{ref:Arnesen2001}. In
this context, the existence of quantum correlations have been put
in connection with phase transitions
\cite{ref:Osterloh2002,ref:Osborne2002}. Thermal entanglement has
been studied in spin chains described by Heisenberg models
\cite{ref:Gong2009}, in atom-cavity systems \cite{ref:Wang2009},
in simple molecular models \cite{ref:Pal2010}, and has been
proposed as a resource in quantum teleportation protocols
\cite{ref:Zhou2009}. Nonclassical and nonlocal correlations in
thermalized quantum systems have been investigated
\cite{ref:Werlang2010, ref:Souza2009}.

Thermal entanglement has been studied in spin-star networks. For
instance, Hutton and Bose \cite{ref:Hutton2004} have analyzed the
zero-temperature properties of such quantum systems, bringing to
light interesting properties related to the parity of the number
of outer (peripheral) spins. Recently, Wan-Li {\it et al} have
studied the thermal entanglement in a spin-star network with three
peripheral spins \cite{ref:Wan-Li2009}, evaluating pairwise
entanglement between all possible couples of spins. More recently,
Anz\`{a} {\it et al} \cite{ref:Anza2010} have analyzed tripartite
correlations in a similar system, exploiting the tripartite
negativity. Nevertheless, as already pointed out, such a tool
cannot distinguish between tripartite entanglement and generalized
biseparability.

In this paper, we investigate tripartite entanglement in the same
system analyzed by Anz\`{a} {\it et al}, but exploiting the new
criteria introduced by Huber {\it et al}. To this end, in the next
section we summarize the results of ref \cite{ref:Huber2010} and
specialize them to the three-spin case. In the third section we
apply these tools to a thermalized spin-star network made of three
peripheral spins interacting with a central one, bringing to light
the presence of genuine tripartite thermal entanglement. Finally,
in the last section, we discuss our results and give some
conclusive remarks.

\section{Detection of Tripartite Entanglement}

In a recent paper by Huber {\it et al} \cite{ref:Huber2010}, it
has been shown that given a biseparable density operator $\rho$
acting on the Hilbert space $\cal H$, whether corresponding to
pure or mixed state, for any completely separable state
$\Ket{\Psi}$ of the {\it duplicated} Hilbert space ${\cal
H}\otimes{\cal H}$, it turns out that
\begin{eqnarray}
\nonumber &&{\cal Q}(\rho, \Psi) = \sqrt{\Bra{\Psi}\,\rho^{\otimes
2}\,\Pi\,\Ket{\Psi}}\\
\nonumber && \,\, - \sum_i \sqrt{\Bra{\Psi}\,(\Pi_{A_i}\otimes
\mathbf{1}_{B_i})^\dag\, \rho^{\otimes
2}\,(\Pi_{A_i}\otimes \mathbf{1}_{B_i})\,\Ket{\Psi}} \le 0\,,\\
\label{eq:General_Definition_Of_Q}
\end{eqnarray}
where $i$ runs over all possible bipartitions of the system. The
operator $\Pi$ performs swapping between the two parts of the
duplicated Hilbert space, in the following way:
\begin{equation}
  \Ket{\psi_1}\otimes\Ket{\psi_2}\,\epsilon\, {\cal H}\otimes{\cal H}
  \Rightarrow \Pi\Ket{\psi_1}\otimes\Ket{\psi_2} =
  \Ket{\psi_2}\otimes\Ket{\psi_1}\,.
\end{equation}
Moreover, for a bipartition of the system $(A_i, B_i)$ and any
separable state
$\Ket{\Psi}=\Ket{\psi_{A_i}}\otimes\Ket{\psi_{B_i}}\otimes\Ket{\chi_{A_i}}\otimes\Ket{\chi_{B_i}}\,\epsilon\,$
${\cal H}\otimes{\cal H}$, one has:
\begin{equation}
  (\Pi_{A_i} \otimes \mathbf{1}_{B_i}) \Ket{\Psi} =
  \Ket{\chi_{A_i}}\otimes\Ket{\psi_{B_i}}\otimes\Ket{\psi_{A_i}}\otimes\Ket{\chi_{B_i}}\,.
\end{equation}

On the basis of \eqref{eq:General_Definition_Of_Q}, the occurrence
of the condition ${\cal Q}
> 0$ for some trial state $\Ket{\Psi}$ guarantees that state $\rho$ possesses genuine multipartite
entanglement, in the sense that it is neither simply biseparable
nor biseparable in a generalized sense (i.e., a state of the form
$\rho=\sum_i\,p_i\,\rho_{A_i} \otimes \rho_{B_i}$). Therefore,
after introducing the positive part of ${\cal Q}$,
\begin{equation}
{\cal C}(\rho, \Psi)=\max[0, {\cal Q}(\rho, \Psi)]\,,
\end{equation}
for a finite-dimensional Hilbert space, we can use the following
condition,
\begin{equation}
  {\cal I}(\rho) = \int d\Psi\,{\cal C}(\rho, \Psi) > 0\,, \label{eq:GeneralSufficientCondition}
\end{equation}
as a sufficient condition to assert that $\rho$ is a multipartite
${\it entangled}$ state. In \eqref{eq:GeneralSufficientCondition}
the integration is meant over all possible completely separable
states of the ${\cal H}\otimes{\cal H}$ Hilbert space. This means
that, if the state $\Ket{\Psi}$ depends on $P$ parameters, $q_1$,
... $q_P$, one has $d\Psi = \prod_{k=1}^{P} \mathrm{d} q_k$.

Let us specialize this analysis to a three-spin system. In order
to afford numerical calculation, we prevent integration over all
possible {\it trial} states, and we consider only the special case
\begin{eqnarray}
\nonumber\Ket{\Psi(\theta,\phi,\eta,\xi)}&=&\Ket{\theta,\phi}\otimes\Ket{\theta,\phi}\otimes\Ket{\theta,\phi}\\
&\otimes&\Ket{\eta,\xi}\otimes\Ket{\eta,\xi}\otimes\Ket{\eta,\xi}\,,\label{eq:TrialState}
\end{eqnarray}
where
\begin{equation}
\Ket{\alpha,\beta}=\cos\alpha\Ket{0}+e^{-i\beta}\sin\alpha\Ket{1}\,.
\end{equation}
This choice gives rise to the following expression for the
quantity ${\cal Q}$:

\begin{widetext}
\begin{eqnarray}
\nonumber {\cal Q} (\rho, \theta, \phi, \eta, \xi) &=& |\Bra{\theta,\phi}\Bra{\theta,\phi}\Bra{\theta,\phi}{\rho}\Ket{\eta,\xi}\Ket{\eta,\xi}\Ket{\eta,\xi}|\\
\nonumber &-&\sqrt{\Bra{\eta,\xi}\Bra{\eta,\xi}\Bra{\theta,\phi}{\rho}\Ket{\theta,\phi}\Ket{\eta,\xi}\Ket{\eta,\xi}\,\Bra{\theta,\phi}\Bra{\theta,\phi}\Bra{\eta,\xi}{\rho}\Ket{\eta,\xi}\Ket{\theta,\phi}\Ket{\theta,\phi}}\\
\nonumber &-&\sqrt{\Bra{\theta,\phi}\Bra{\eta,\xi}\Bra{\eta,\xi}{\rho}\Ket{\eta,\xi}\Ket{\eta,\xi}\Ket{\theta,\phi}\,\Bra{\eta,\xi}\Bra{\theta,\phi}\Bra{\theta,\phi}{\rho}\Ket{\theta,\phi}\Ket{\theta,\phi}\Ket{\eta,\xi}}\\
&-&\sqrt{\Bra{\eta,\xi}\Bra{\theta,\phi}\Bra{\eta,\xi}{\rho}\Ket{\eta,\xi}\Ket{\theta,\phi}\Ket{\eta,\xi}\,\Bra{\theta,\phi}\Bra{\eta,\xi}\Bra{\theta,\phi}{\rho}\Ket{\theta,\phi}\Ket{\eta,\xi}\Ket{\theta,\phi}}\,,
\end{eqnarray}
\end{widetext}
where the first {\it bra} in a product
$\Bra{\psi_3}\Bra{\psi_2}\Bra{\psi_1}$ refers to the third spin,
and so on. The relevant positive part is:
\begin{eqnarray}
{\cal C} (\rho, \theta, \phi, \eta, \xi) &=& \max\left[0,\, {\cal
Q} (\rho, \theta, \phi, \eta, \xi)\right]\,.
\end{eqnarray}

In order to further simplify the calculation associated to
\eqref{eq:GeneralSufficientCondition}, we introduce the following
nonnegative quantity:
\begin{widetext}
\begin{eqnarray}
{\cal I}^{(N)} (\rho) = \sum_{j=0}^{N-1}\sum_{k=0}^{N-1}\int_0^\pi
d\theta\,\int_0^\pi d\eta\, {\cal C}(\rho,\,\theta,\,2\pi j / N,\,
\eta,\, 2\pi k / N). \label{eq:IntegralDetector}
\end{eqnarray}
\end{widetext}
where integration over the longitudinal angles, $\phi$ and $\xi$,
has been replaced by a finite sum.

\begin{figure}
\subfigure[]{\includegraphics[width=0.40\textwidth,
angle=0,clip=a]{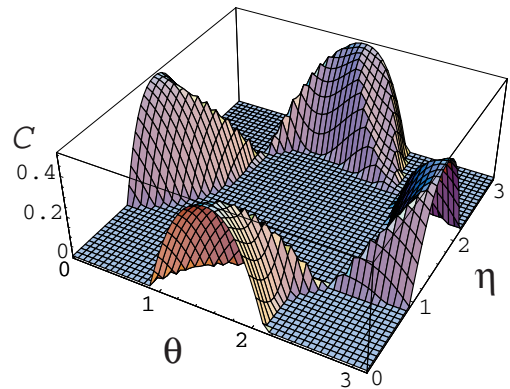}} %
\subfigure[]{\includegraphics[width=0.40\textwidth,
angle=0,clip=b ]{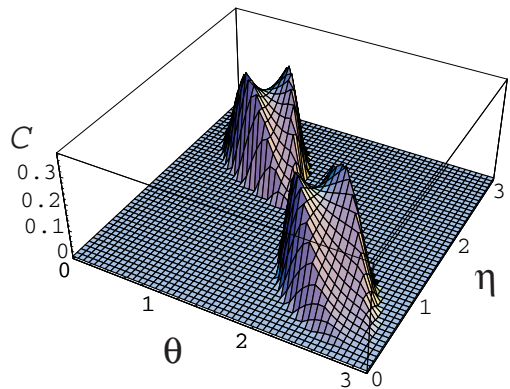}}%
\caption{(Color online). The function ${\cal C}(\rho, \theta, 0,
\eta, 0)$ for the $GHZ$-state (a) and for the $W$-state (b).}
\label{fig:CFunction_WState_GHZ}
\end{figure}

For any biseparable state $\rho$ one has ${\cal I}^{(N)}
(\rho)=0$, and then, conversely, strict positivity of such
quantity is a sufficient condition for the state $\rho$ to be a
genuinely tripartite entangled state. Though this condition is not
strong as \eqref{eq:GeneralSufficientCondition}, we will prove
that it allows revealing of tripartite entanglement in an
effective way. In particular, in
Fig.~\ref{fig:CFunction_WState_GHZ} it is shown the function
${\cal C}(\rho, \theta, 0, \eta, 0)$ for the two archetypical
tripartite entangled states:
$\Ket{GHZ}=(\Ket{000}+\Ket{111})/\sqrt{2}$ and
$\Ket{W}=(\Ket{100}+\Ket{010}+\Ket{001})/\sqrt{3}$. The analytical
expression of ${\cal C}(\rho, \theta, 0, \eta, 0)$ can be easily
given for these two states:

\begin{subequations}
\begin{eqnarray}
  \nonumber
  &&{\cal C}(\rho_W, \theta, 0, \eta, 0)  = 3
  \left|\cos\theta\,\cos\eta\right|\,\sin^2\theta\,\sin^2\eta\\
  \nonumber
  && \qquad -\left|\sin\theta\,\sin\eta\, (2\cos\theta\,\sin\eta+\sin\theta\,\cos\eta)\right|\\
  && \qquad
  \times\left|(2\cos\eta\,\sin\theta+\sin\eta\,\cos\theta)\right|\,,
\end{eqnarray}
and
\begin{eqnarray}
  \nonumber
  &&{\cal C}(\rho_{GHZ}, \theta, 0, \eta, 0) =
  \frac{3}{2}|\cos^2\eta\,\cos\theta+\sin^2\eta\,\sin\theta|\\
  \nonumber
  && \qquad \times|\cos^2\theta\,\cos\eta+\sin^2\theta\,\sin\eta|\\
  \nonumber
  && \qquad + |3\cos\eta+\cos 3\eta + 4\sin^3\eta|\\
  && \qquad \times |3\cos\theta+\cos 3\theta +
  4\sin^3\theta|/32\,,
\end{eqnarray}
\end{subequations}
where $\rho_W=\SimKet{W}\SimBra{W}$ and
$\rho_{GHZ}=\SimKet{GHZ}\SimBra{GHZ}$. It is well visible that
${\cal C}$ is far from being identically vanishing, for these two
states.

The analytical calculation of the same quantity for the state
$\SimKet{\tilde{W}}=(\Ket{110}+\Ket{101}+\Ket{011})/\sqrt{3}$ can
be easily carried on, and it gives a result very similar to that
obtained for the $W$-state, provided the swapping of all the
trigonometric functions: $\sin\leftrightarrows\cos$. Moreover, we
have performed the same analysis for separable states of different
kinds, and we have always found that the corresponding ${\cal
C}$-function is zero everywhere. Integration of the functions
plotted in Fig.~\ref{fig:CFunction_WState_GHZ} provides ${\cal
I}^{(1)}$ for the two states. Performing the integration over
$\theta$ and $\eta$ with a $15\times 15$, grid we have got ${\cal
I}^{(1)}(\rho_W)\approx 0.36$ and ${\cal
I}^{(1)}(\rho_{GHZ})\approx 0.75$, while, spanning over four
remarkable longitudinal angles ($0$, $\pi/2$, $\pi$, $3\pi/2$), we
have got ${\cal I}^{(4)}(\rho_W)\approx 2.87$ and ${\cal
I}^{(4)}(\rho_{GHZ})\approx 11.46$.

\begin{figure}
\subfigure[]{\includegraphics[width=0.40\textwidth,
angle=0,clip=a]{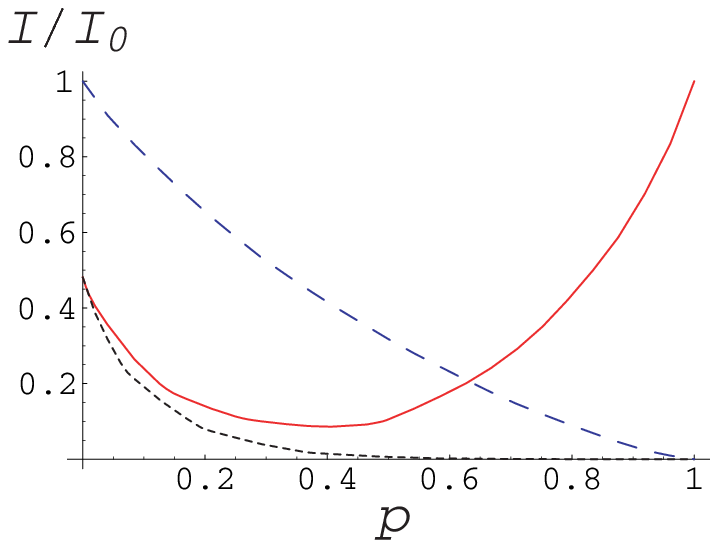}} %
\subfigure[]{\includegraphics[width=0.40\textwidth,
angle=0,clip=b ]{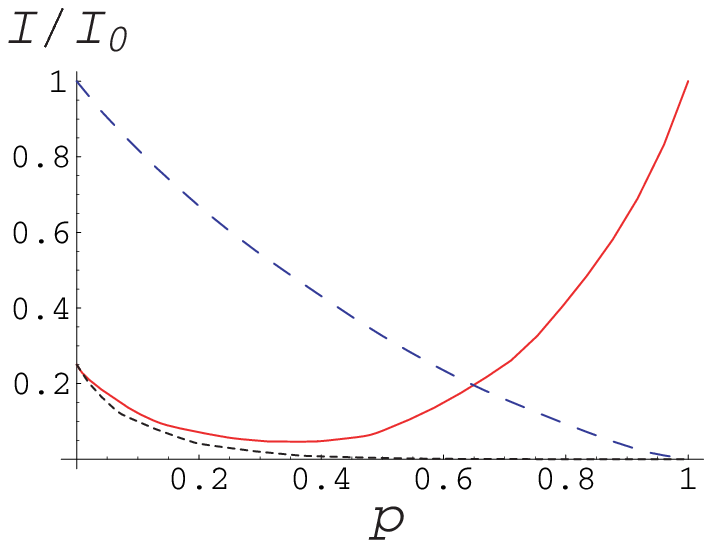}}%
\caption{(Color online). The functions ${\cal I}^{(N)}(\rho)/{\cal
I}^{(N)}_0$ for three exemplar mixed states. Figure (a)
corresponds to $N=1$ and figure (b) to $N=4$. In both figures, the
solid (red) line corresponds to the state
$\rho=p\Projector{GHZ}+(1-p)\Projector{W}$, the dotted (black)
line to the state $\rho=p\Projector{111}+(1-p)\Projector{W}$, and
the dashed (blue) line to the state
$\rho=p\Projector{111}+(1-p)\Projector{GHZ}$. }
\label{fig:Detection_MixedStates}
\end{figure}

In Fig.~\ref{fig:Detection_MixedStates} we show the function
${\cal I}^{(N)}(\rho)$ for three classes of mixed states: mixtures
of $\Ket{GHZ}$ and $\Ket{W}$, mixtures of $\Ket{GHZ}$ and the
factorized state $\Ket{111}$, and mixtures of $\Ket{W}$ and
$\Ket{111}$. In this figure and in the next analogous ones, we
plot the ratios between ${\cal I}^{(N)}(\rho)$ and ${\cal
I}^{(N)}_0 = {\cal I}^{(N)}(\rho_{GHZ})$. It is well visible that
${\cal I}^{(1)}(\rho)$ and ${\cal I}^{(4)}(\rho)$ approach zero as
the state approaches a factorized state, while these quantities
reach higher values as the state possesses tripartite
entanglement.

\begin{figure}
\includegraphics[width=0.40\textwidth,
angle=0]{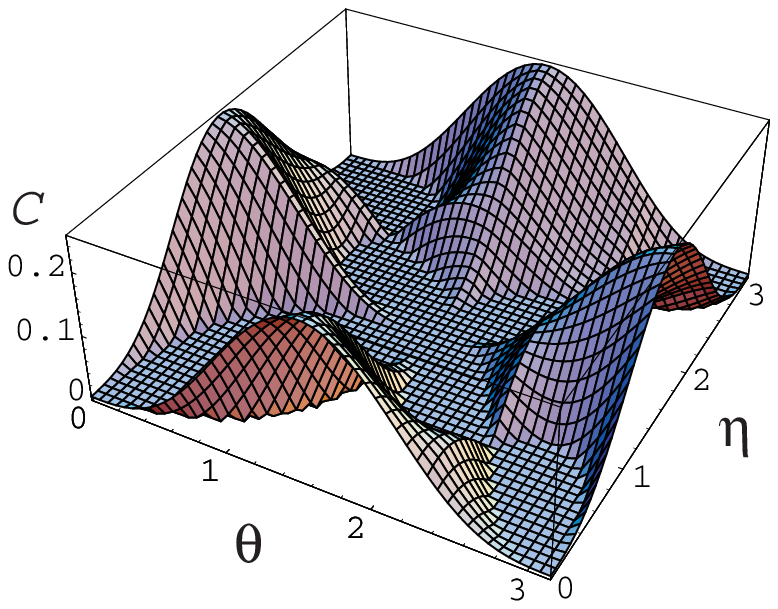} %
\caption{(Color online). The function ${\cal C}(\rho, \Psi)$ for
the $GHZ$-like state $\Ket{\sigma
GHZ}=(\Ket{110}+\Ket{001})/\sqrt{2}$ with the trial state
$\Ket{\theta,0}\otimes\Ket{\theta,0}\otimes\Ket{\eta,0}
\otimes\Ket{\eta,0}\otimes\Ket{\eta,0}\otimes\Ket{\theta,0}$. }
\label{fig:CFunction_SimGHZ}
\end{figure}

This analysis supports the idea that the criteria introduced in
\cite{ref:Huber2010} are quite effective in revealing genuine
tripartite entanglement. Nevertheless, it is important to note
that the subset of trial states considered plays a very
fundamental role in the detection of multipartite entanglement.
Indeed, if we consider for example trial states of the form
$|\tilde{\Psi}\rangle=\Ket{\theta,\phi}\otimes\Ket{\theta,\phi}\otimes\Ket{\eta,\xi}
\otimes\Ket{\eta,\xi}\otimes\Ket{\eta,\xi}\otimes\Ket{\theta,\phi}$,
then we are not able to detect entanglement of the GHZ-state. On
the contrary, this choice is able to detect tripartite
entanglement of the state $\Ket{\sigma
GHZ}=(\Ket{110}+\Ket{001})/\sqrt{2}$, which instead never violates
the inequality ${\cal Q}\le 0$ when the trial state has the form
given in \eqref{eq:TrialState}. In fact, on the one hand, it is
${\cal I}^{(1)}(\rho_{\sigma GHZ})={\cal I}^{(4)}(\rho_{\sigma
GHZ}) = 0$, while on the other hand, in Fig.
\ref{fig:CFunction_SimGHZ} we can see that, when the trial state
has the form $|\tilde{\Psi}\rangle$, it turns out to be ${\cal Q}
> 0$ in a wide range.

In spite of these limitations related to spanning a subset of the
relevant Hilbert space, we will use the functionals ${\cal
I}^{(N)}(\rho)$ defined in \eqref{eq:IntegralDetector} to carry on
our analysis, both for the sake of simplicity and since we think
it is effective enough in our problem.

\section{Thermal Tripartite Entanglement}

Spin-star networks have been studied in connection with
decoherence problems, especially in the analysis of the
Non-Markovian character of spin baths \cite{ref:SSN-Decoherence},
and for applications in quantum information
\cite{ref:SSN-QuantumInfo}.

In a recent paper, Wan-Li {\it et al} \cite{ref:Wan-Li2009} have
studied the thermal entanglement in a spin-star system made of a
central spin coupled to three peripheral spins through an
anisotropic $\sigma$-$\sigma$ interactions (the longitudinal
(\lq$\sigma_z$-$\sigma_z$\rq) interaction and the total transverse
interaction (\lq$\sigma_x$-$\sigma_x$\rq $+$
\lq$\sigma_y$-$\sigma_y$\rq) have independent coupling strengths)
identical for the three outer spins. More recently, a similar
system has been studied, removing the longitudinal (i.e.,
$\sigma_z-\sigma_z$) interaction from the coupling between the
spins, and introducing a certain inhomogeneity in the coupling
strengths between the central spin and the outer ones
\cite{ref:Anza2010}. Here we examine the same model, then
considering the following Hamiltonian:
\begin{equation}
  H = \frac{\omega_0}{2}\,\sum_{k=0}^{3}\, \hat{\sigma}_z^{k} \, + \,
  \sum_{k=1}^{3}\, c_k
  (\hat{\sigma}_+^{0}\hat{\sigma}_-^{k}+\hat{\sigma}_-^{0}\hat{\sigma}_+^{k})\,,
  \label{eq:Hamiltonian}
\end{equation}
where $\hat{\sigma}_\alpha^{k}$ is the Pauli operator along the
direction $\alpha$ ($\alpha=x,y,z$) of the spin $k$ ($k=0,1,2,3$),
$\hat{\sigma}_\pm^{k}$ are the corresponding raising and lowering
operators, $\omega_0$ is the free Bohr frequency of all the spins
due to an external magnetic field, and $c_k$ is the coupling
constant between the spin $0$ and the $k$-th one.

Once the system reaches the thermodynamical equilibrium, it can be
described by the thermal state,
\begin{equation}
  \rho^{(\mathrm{T})} = \frac{e^{- H / kT}}{\trace(e^{- H / kT})}\,\,,
\end{equation}
which has the same eigenstates of the Hamiltonian $H$. The result
of the diagonalization of $H$ is reported in the Appendix
\ref{app:Diagonalization}.

In ref \cite{ref:Anza2010}, Anz\`{a} {\it et al} have considered
the homogeneous case ($c_1=c_2=c_3=c$) and different kinds of
inhomogeneous models. In the following we will consider both the
homogeneous model and the inhomogeneous case $c_1=c_3=c$ and
$c_2=c\,x$, with $x$ a dimensionless inhomogeneity parameter. We
will apply the new criteria for multipartite entanglement
detection to the state obtained starting from the four-qubit
thermal state and tracing over the degrees of freedom of the
central spin:
\begin{equation}
  \rho^{(\mathrm{P})} = \trace_0(\rho^{(\mathrm{T})})\,\,,
\end{equation}
which describes the three peripheral spins.

\subsection{Homogeneous Model}

The homogeneous model has been studied by Wan-Li {\it et al}
\cite{ref:Wan-Li2009} (with the addition of a longitudinal
coupling) and by Anz\`{a} {\it et al} \cite{ref:Anza2010}. In the
first paper, the pairwise entanglement has been studied, through
the use of concurrences. In the second paper, tripartite
correlations have been investigated, through the use of the
tripartite negativity \cite{ref:Sabin2008}.

Tripartite negativity is an imperfect tool to detect genuine
tripartite entanglement, since it cannot distinguish between this
form of entanglement and generalized biseparability. Nevertheless,
it has helped to find points wherein tripartite correlations are
significant, even if to disclose the nature of these correlations
one needs a further analysis.

On the basis of the criteria proposed in ref \cite{ref:Huber2010},
it is possible to assert in an indisputable way the presence of
tripartite entanglement when the condition ${\cal Q} > 0$ is
fulfilled. The quantity in \eqref{eq:GeneralSufficientCondition}
and its simplified version in \eqref{eq:IntegralDetector}, provide
sufficient conditions for the presence of tripartite entanglement.
Moreover, one could think that they furnish sorts of degree of
entanglement, in the sense that higher values of these quantities
can be understood as higher or wider violations of the inequality
${\cal Q} \le 0$. Notwithstanding, it is important to stress that
neither ${\cal I}^{(N)}$ nor ${\cal I}$ provide a measure of
entanglement, and that in the case of ${\cal I}^{(N)}$ there is
also the problem that a limited part of the relevant Hilbert space
is spanned in the integration process, as already pointed out.

\begin{figure}
\includegraphics[width=0.40\textwidth,
angle=0]{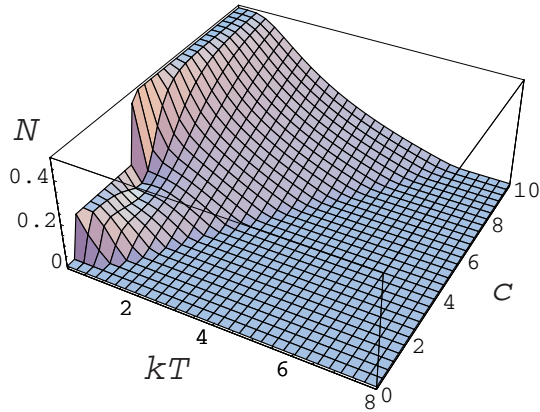} %
\caption{(Color online). The negativity for the state of the
peripheral system ${\cal N}(\rho^{(\mathrm{P})})$ vs temperature
($kT$ in units of $\omega_0^{-1}$) and coupling constant ($c$ in
units of $\omega_0^{-1}$).
%Integration over $\theta$ and $\eta$ is performed with a $15\times 15$ grid.
} \label{fig:Negativity_Thermal_vs_T}
\end{figure}

\begin{figure}
\includegraphics[width=0.40\textwidth,
angle=0]{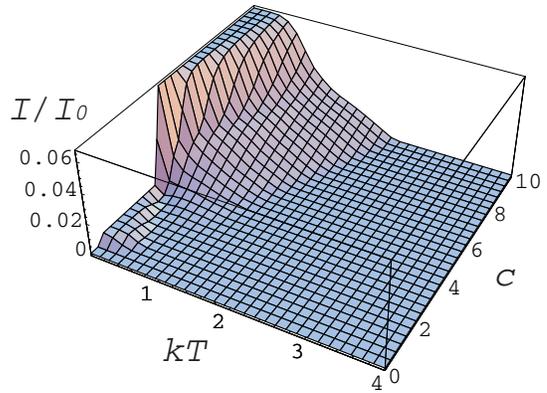} %
\caption{(Color online). The function ${\cal
I}^{(1)}(\rho^{(\mathrm{P})})/{\cal I}^{(1)}_0$ for the state of
the peripheral system vs temperature ($kT$ in units of
$\omega_0^{-1}$) and coupling constant ($c$ in units of
$\omega_0^{-1}$).
%Integration over $\theta$ and $\eta$ is performed with a $15\times 15$ grid.
}
\label{fig:Detection_Thermal_vs_T}
\end{figure}

\begin{figure}
\subfigure[]{\includegraphics[width=0.40\textwidth,
angle=0,clip=a]{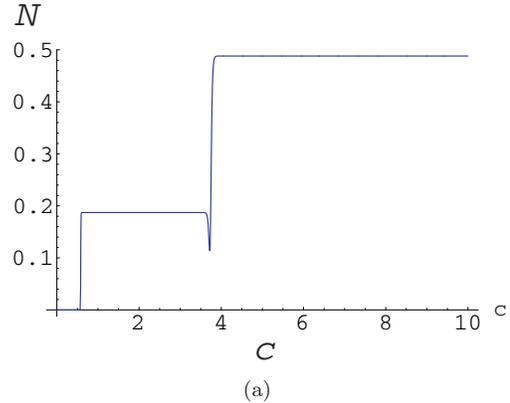}} %
\subfigure[]{\includegraphics[width=0.40\textwidth,
angle=0,clip=b ]{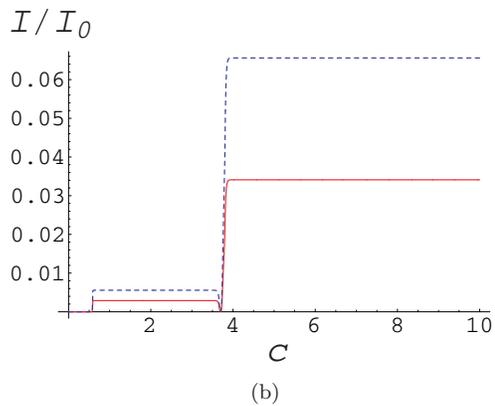}}%
\caption{Figure (a) shows the low temperature profile of the
tripartite negativity of the peripheral state ${\cal
N}(\rho^{(\mathrm{P})})$ versus the coupling constant $c$ (in
units of $\omega_0^{-1}$). Figure (b) shows the low temperature
profiles of ${\cal I}^{(1)}(\rho^{(\mathrm{P})})/{\cal I}^{(1)}_0$
(solid curve, red online) and ${\cal
I}^{(4)}(\rho^{(\mathrm{P})})/{\cal I}^{(4)}_0$ (dashed curve,
blue online) . In both figures, the temperature is such that
$kT/\omega_0=0.01$..
%Integration over $\theta$ and $\eta$ is performed with a $15\times 15$ grid.
}
\label{fig:Detection_Thermal_Zero}
\end{figure}

Fig. \ref{fig:Negativity_Thermal_vs_T} and
\ref{fig:Detection_Thermal_vs_T} show the tripartite negativity
${\cal N}(\rho^{(\mathrm{P})})$ and the quantity ${\cal
I}^{(1)}(\rho^{(\mathrm{P})})$, respectively, as functions of both
the temperature and the coupling constant between the central spin
and the peripheral ones. Fig. \ref{fig:Detection_Thermal_Zero}
shows the quantities ${\cal I}^{(N)}(\rho^{(\mathrm{P})})$, for
$N=1$ and $N=4$, as functions of the coupling constant, at low
temperature. The behaviors are qualitatively very similar: for
increasing temperature the quantity ${\cal
I}^{(1)}(\rho^{(\mathrm{P})})$ decreases, while at very low
temperature abrupt changes are well visible at specific values of
the coupling constant. In particular, for $kT/\omega_0=0.01$,
around the value of the coupling constant $c=0.6\omega_0$, there
is a first transition from $0$ to a positive value, and around
$c=3.7\omega_0$ another transition is well visible. These
transitions, revealed by all the {\it witness} quantities here
considered, correspond to very abrupt changes of the ground state
of the four-qubit system. In particular, for $c<0.6\omega_0$ the
ground state is $\Ket{\psi_8}=\Ket{0000}$, for
$0.6\omega_0<c<3.7\omega_0$ the lowest energy state is
$\Ket{\psi_4^-}$, and for $c>3.7\omega_0$ the ground state is
$\Ket{\psi_2^-}$ (see Appendix \ref{app:Diagonalization} for the
explicit expression of these states). The corresponding
three-qubit states are:
$\rho^{(\mathrm{P})}\approx\rho^{(8)}=\Ket{000}\Bra{000}$,
$\rho^{(\mathrm{P})}\approx\rho^{(4-)}=0.5\,\Ket{111}\Bra{111}+0.5\,\SimKet{\tilde{W}}\SimBra{\tilde{W}}$,
and
$\rho^{(\mathrm{P})}\approx\rho^{(2-)}=0.5\,\SimKet{W}\SimBra{W}+0.5\,\SimKet{\tilde{W}}\SimBra{\tilde{W}}$,
respectively.

\subsection{Inhomogeneous Model}

\begin{figure}
\includegraphics[width=0.40\textwidth,
angle=0]{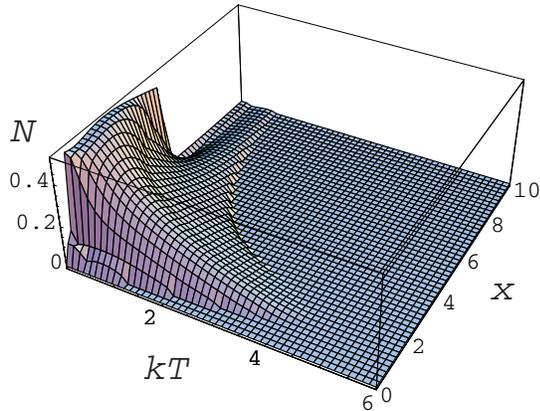} %
\caption{(Color online). The tripartite negativity of the
peripheral state, ${\cal N}(\rho^{(\mathrm{P})})$, for the
inhomogeneous model, as a function of both temperature ($kT$ in
units of $\omega_0^{-1}$) and inhomogeneity parameter, with
$c=6\omega_0$.} \label{fig:Negativity_Inhomogeneous}
\end{figure}

\begin{figure}
\subfigure[]{\includegraphics[width=0.40\textwidth,
angle=0,clip=a]{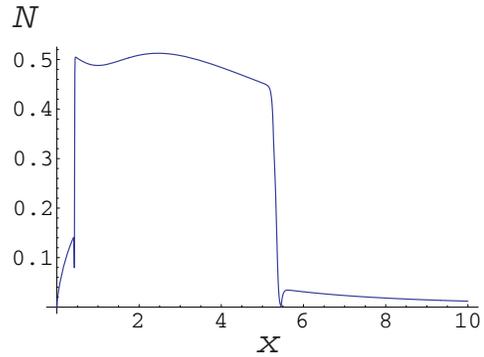}} %
\subfigure[]{\includegraphics[width=0.40\textwidth,
angle=0,clip=b ]{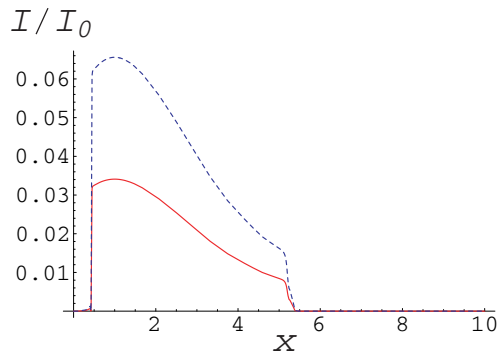}}%
\caption{(Color online). Figure (a) shows the low temperature
profile of the tripartite negativity of the peripheral state
${\cal N}(\rho^{(\mathrm{P})})$ versus the inhomogeneity parameter
$x$. Figure (b) shows the low temperature profiles of ${\cal
I}^{(1)}(\rho^{(\mathrm{P})})/{\cal I}^{(1)}_0$ (solid curve, red
online) and ${\cal I}^{(4)}(\rho^{(\mathrm{P})})/{\cal I}^{(4)}_0$
(dashed curve, blue online) . In both figures, the coupling
constant is $c=6\omega_0$ and the temperature is such that
$kT/\omega_0=0.01$.} \label{fig:Detector_vs_Negativity}
\end{figure}

In \cite{ref:Anza2010}, it has also been analyzed the effect of
anisotropy in the coupling constants. The analysis based on the
tripartite negativity shows that, in spite of the lack of symmetry
of the system, the degree of correlation between the three
peripheral spins can still be appreciable. In particular, it has
been brought to the light the fact that at low temperature the
maximum of tripartite negativity is reached for values of the
inhomogeneity parameter different from (larger than) unity. Such
behavior is well visible in Fig.
\ref{fig:Negativity_Inhomogeneous} and Fig.
\ref{fig:Detector_vs_Negativity}a. This unexpected result
seemingly suggests that the maximum of tripartite correlations
does not correspond to the maximum of symmetry of the system.

\begin{figure}
\includegraphics[width=0.40\textwidth,
angle=0]{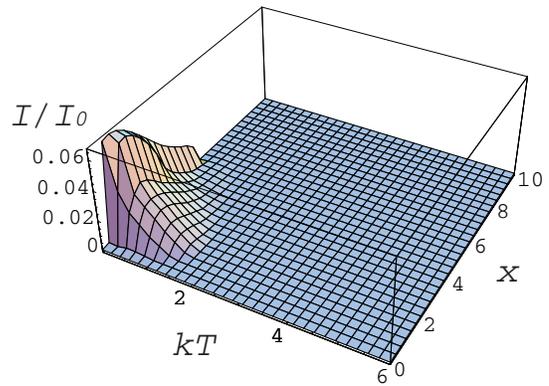} %
\caption{(Color online). The function ${\cal
I}^{(1)}(\rho^{(\mathrm{P})})/{\cal I}^{(1)}_0$ for the
thermalized system in the presence of inhomogeneity, for
$c=6\omega_0$. This quantity is plotted versus the temperature
($kT$ in units of $\omega_0^{-1}$) and the inhomogeneity parameter
$x$.
%Integration over $\theta$ and $\eta$ is performed with a $15\times 15$ grid.
} \label{fig:Detection_Thermal_Inhomogeneous}
\end{figure}

It can be interesting to compare such results with those coming
from the tools based on the work by Huber {\it et al}
\cite{ref:Huber2010}. Fig.
%\ref{fig:Detection_Thermal_Inhomogeneous}a
\ref{fig:Detection_Thermal_Inhomogeneous} shows the quantity
${\cal I}^{(1)}(\rho^{(\mathrm{P})})$ as a function of temperature
and anisotropy parameter $x$, for $c=6\omega_0$. Fig.
\ref{fig:Detector_vs_Negativity} shows the low temperature
profiles, where fast transitions are very well visible. The local
maximum of the tripartite negativity around $x=2.5$ is
appreciable. On the contrary, the quantity ${\cal
I}^{(4)}(\rho^{(\mathrm{P})})$ does not exhibit the same behavior.
Instead, it does possess a maximum in $x=1$.

At low temperature, both ${\cal N}(\rho^{(\mathrm{P})})$ and the
${\cal I}^{(N)}$ functionals have significant values for
intermediate values of $x$, say for $0.5\lesssim x \lesssim 5$,
and are small or vanishing out of this region. Note that for very
small $x$ ($c_2 \ll c_1, c_3$) one of the spins is almost
uncoupled to the central one. On the other hand, for large $x$
($c_2 \gg c_1, c_3$) that spin has a much stronger coupling
constant than the other two, whose couplings can then be
considered as a perturbation, so that, at zeroth order, the latter
two spins are uncoupled to the central one. In both cases, it is
physically reasonable that tripartite correlations between the
outer spins are negligible or absent. The abrupt changes of the
{\it witness} quantities are related to the sudden modifications
of the ground state of the system. However, we remark that low or
vanishing values of the tripartite negativity and ${\cal I}^{(N)}$
(or even ${\cal I}$) do not guarantee the absence of tripartite
entanglement or tripartite correlations. Conversely, the
non-vanishing values of some ${\cal I}^{(N)}$ functionals
guarantee the presence of tripartite entanglement.

\section{Discussion} %{Conclusive Remarks}

In this paper we have investigated the tripartite thermal
entanglement in a spin-star network with three peripheral spins.
The interaction with the central spin is responsible for the
establishment of tripartite correlations between the peripheral
ones, and such correlations survive even when the system is at
thermal equilibrium. We have considered both the homogeneous
model, where all the coupling constants are equal, and the
inhomogeneous model, where one of the outer spins is coupled to
the central one with a different strength.

The analysis is carried on through the use of the quantities
defined in \eqref{eq:IntegralDetector}, which is a simplified
version of \eqref{eq:GeneralSufficientCondition}, where a limited
region of the relevant Hilbert space is spanned. Each of these
functionals has the property that its strict positivity guarantees
the presence of genuine tripartite entanglement. Nevertheless, it
must be clarified that none of such quantities provides a measure
of the amount of tripartite entanglement. Anyway, a larger value
of ${\cal I}$ does mean a higher or wider violations of the
condition ${\cal Q} \le 0$. Therefore, one can conjecture that a
higher value of ${\cal I}$ corresponds to a state that exhibits
entanglement more than other states. The same assertion is weaker
when applied to ${\cal I}^{(N)}$, since evaluation of this
functional does not require spanning over all of the Hilbert
space. Moreover, it is important to know that the use of different
${\cal I}$-functionals (${\cal I}$, ${\cal I}^{(N)}$ with
different $N$, or other similar quantities that consider spanning
on different subsets of the relevant Hilbert space) could lead to
different predictions.

For the homogeneous model, the low temperature behavior is
characterized by abrupt changes of the quantities ${\cal I}^{(N)}$
versus the coupling constant. These {\it transitions} correspond
to concomitant abrupt changes of the system ground state. At
higher temperature, ${\cal I}^{(1)}$ goes to zero. For the
inhomogeneous model, the dependence of ${\cal I}^{(1)}$ on the
inhomogeneity parameter and temperature, when the coupling
constant is fixed at some high value, is again characterized by
abrupt changes with respect to $x$ at low temperature, and by
vanishing at high temperature. It is remarkable that, at low
temperature, the dependence on the inhomogeneity parameter reveals
the presence of a maximum for $x=1$, i.e. in the homogeneous case.
This result, on the one hand is seemingly in line with
expectations coming from intuition, and on the other hand is
supposedly different from the predictions coming from the use of
tripartite negativity. Nevertheless, different behaviors of these
quantities do not imply contradictions, since neither tripartite
negativity nor the ${\cal I}$-quantities provide necessary
conditions for the presence of tripartite entanglement or a
measure of such form of entanglement. What is sure is that in the
parameter region where ${\cal I}^{(N)}$ is non vanishing, the
thermal state possesses genuine tripartite entanglement.

Therefore, in spite of the limitations of our analysis, we have
found genuine tripartite entanglement in our system at thermal
equilibrium, even at non vanishing temperature and in the presence
of inhomogeneity.

\appendix

\section{Diagonalization of $H$} \label{app:Diagonalization}

In this appendix we give eigenvalues and eigenvectors of the
Hamiltonian in \eqref{eq:Hamiltonian}, as functions of $x$, $c>0$
and $\omega_0$. The homogeneous model is obtained for $x=1$.

The eigenvalues of the Hamiltonian are:
\begin{subequations}
\begin{eqnarray}
E_1^{\pm} &=& \pm c\,,\\
E_2^{\pm} &=& \pm \frac{c}{2} \left[x +
\left(8+x^2\right)^{\frac{1}{2}}\right]\,,\\
E_3^{\pm} &=& \pm \frac{c}{2}
\left[x-\left(8+x^2\right)^{\frac{1}{2}} \right]\,, \\
E_4^{\pm} &=& \pm \left[c
\left(2+x^2\right)^{\frac{1}{2}}+\omega_0
\right]\,,\\
E_5^{\pm} &=& \pm \left[c \left(2+x^2\right)^{\frac{1}{2}} -
\omega_0 \right]\,,\\
E_6 &=& -\omega_0\,, \\
E_7 &=& \omega_0\,,\\
E_8 &=& -2\omega_0\,,\\
E_9 &=& 2\omega_0\,,
\end{eqnarray}
\end{subequations}
where the eigenvalues $E_6$ and $E_7$ are twofold degenerate
eigenvalues.

The relevant eigenstates are:
%\begin{widetext}
\begin{subequations}
\begin{eqnarray}
\nonumber
\Ket{\psi_{1}^{\pm}} &=& \frac{1}{2} \left[ \left(\Ket{0011} \pm \Ket{1100}\right) - \left(\Ket{0110} \pm \Ket{1001}\right) \right]\,,\\
\,
\end{eqnarray}
\begin{eqnarray}
\nonumber \Ket{\psi_{2}^{\pm}} &=& \frac{1}{K_1}
\: [ \left(\Ket{0011} \pm \Ket{1100}\right) + \left(\Ket{0110} \pm
\Ket{1001}\right)\\
&+& \frac{\sqrt{8+x^2} - x}{2} \left( \Ket{0101} \pm \Ket{1010}
\right) ]\,,
\end{eqnarray}
\begin{eqnarray}
\nonumber \Ket{\psi_{3}^{\pm}} &=& \frac{1}{K_1} \: [
\left(\Ket{0011} \pm \Ket{1100}\right) + \left(\Ket{0110} \pm
\Ket{1001}
\right)\\
&-& \frac{\sqrt{8+x^2}+x}{2} \left( \Ket{0101} \pm \Ket{1010}
\right) ]\,,
\end{eqnarray}
\begin{eqnarray}
\nonumber \Ket{\psi_{4}^{\pm}} &=& \frac{1}{K_2}
\left[\sqrt{2+x^2}
\Ket{0111}\right.\\
&\pm & \left.\left( \Ket{1011} + x\Ket{1101} + \Ket{1110} \right)
\right]\,,
\end{eqnarray}
\begin{eqnarray}
\nonumber \Ket{\psi_{5}^{\pm}} &=& \frac{1}{K_2} \left[\left(
\Ket{0100} + x
\Ket{0010} + \Ket{0001} \right)\right.\\
&\pm & \left.\sqrt{\left( 2+x^2 \right)} \Ket{1000} \right]\,,
\end{eqnarray}
\begin{eqnarray}
\nonumber
\Ket{\psi_{6}^{\alpha}} &=& \frac{1}{K_3} \left[ \frac{1}{x} \Ket{0001} + \Ket{0010} - \left( \frac{1}{x} + x \right) \Ket{0100} \right] \,,  \\
%\,\\
%\end{eqnarray}
%\begin{eqnarray}
\Ket{\psi_{6}^{\beta}} &=& \frac{1}{\sqrt{1+x^2}} \left(
\Ket{0010} - x \Ket{0001} \right)\,,
\end{eqnarray}

\begin{eqnarray}
\nonumber
\Ket{\psi_{7}^{\alpha}} &=& \frac{1}{K_3} \left[ \frac{1}{x} \Ket{1011} + \Ket{1101} - \left( \frac{1}{x} + x \right) \Ket{1110} \right] \,,  \\
%\,\\
%\end{eqnarray}
%\begin{eqnarray}
\Ket{\psi_{7}^{\beta}} &=& \frac{1}{\sqrt{1+x^2}} \left(
\Ket{1101} - x \Ket{1011} \right)\,,
\end{eqnarray}
\begin{eqnarray}
\Ket{\psi_{8}} &=& \Ket{0000}\,,
\end{eqnarray}
\begin{eqnarray}
\Ket{\psi_9} &=& \Ket{1111}\,,
\end{eqnarray}
\end{subequations}
%\end{widetext}

with

\begin{subequations}
\begin{equation}
{K_1}^2 = 4 + 2 \left( \frac{\sqrt{8+x^2} - x}{2} \right)^2\,,
\end{equation}
\begin{equation}
{K_2}^2 = 2\left(2+x^2\right)\,,
\end{equation}
\begin{equation}
{K_3}^2 = \frac{2}{x^2} + 3 + x^2\,.
\end{equation}
\end{subequations}

\end{document}